\documentclass{article}
\usepackage{pdfpages}
\usepackage{xcolor}
\usepackage{graphicx}
\usepackage{cite}
\usepackage{caption}
\usepackage{subcaption}
\usepackage{authblk}
\usepackage{amsmath}

\usepackage[top=5cm, bottom=5cm, left=3cm, right=2cm, headsep=1.5cm, heightrounded=true]{geometry}
\usepackage{fancyhdr}
\graphicspath{ {{images/}} }

\title{ Holographic  Thermodynamics of BTZ Black Holes   and Tsallis Entropy  }
\author[1,2]{Yahya Ladghami\thanks{ \texttt{yahya.ladghami@ump.ac.ma}}}
\author[1,2]{Brahim Asfour\thanks{ \texttt{brahim.asfour@ump.ac.ma}}}
\author[1,2,3]{Amine Bouali\thanks{ \texttt{a1.bouali@ump.ac.ma}}}
\author[1,2]{Taoufik Ouali\thanks{ \texttt{t.ouali@ump.ac.ma}}}
\author[4,5]{G. Mustafa\thanks{ \texttt{gmustafa3828@gmail.com}}}
\affil[1] {Laboratory of Physics of Matter and Radiation, Mohammed I University, BP 717, Oujda, Morocco}
\affil[2]{Astrophysical and Cosmological Center, BP 717, Oujda, Morocco}
\affil[3]{ Higher School of Education and Training, Mohammed I University, BP 717, Oujda, Morocco}
\affil[4]{ Department of Physics, Zhejiang Normal University, Jinhua 321004, People's Republic of China}
\affil[5]{Research Center of Astrophysics and Cosmology, Khazar University, Baku, AZ1096, 41 Mehseti Street, Azerbaijan}

\begin{document}
	\maketitle
	\begin{abstract}
		
This paper presents a detailed study of the thermodynamics of charged BTZ black holes using the conformal holographic extended thermodynamics formalism and Tsallis statistics. The cornerstone of our thermodynamic framework is the re-scaling of CFT by the conformal factor, which is considered a thermodynamic parameter. Here, the AdS radius is distinct from the CFT radius, allowing for independent variations of the central charge and volume. Our analysis revealed that the thermodynamic behavior of charged BTZ black holes in three dimensions is characterized by the stability and absence of phase transitions, contrasting with the behavior of four-dimensional black holes. The central charge of CFT notably influences the thermal evolution of these black holes, with a smaller central charge leading to faster thermal processes.
 Additionally, the temperature of large black holes is proportional to their entropy. By incorporating Tsallis statistics into our study, we found that the stability of black holes depends on the Tsallis parameter. Black holes are always stable when the Tsallis parameter is less than 2. However, if this parameter is greater than 2, a first-order phase transition occurs between small stable and large unstable. Overall, our findings contribute to a deeper understanding of the holographic thermodynamics of lower-dimensional black holes and the impact of non-extensive statistics on their physical properties.
		
\end{abstract}
\section{Introduction}

Black holes are one of the most significant and crucial areas in theoretical physics. Their importance lies in the potential to find solutions to the most significant challenges in physics, such as the nature of quantum gravity, the unification of the fundamental forces, and the resolution of the information paradox.
	\\
	
	For a long time, it was believed that black holes absorb everything and emit nothing. However, considering quantum effects, Hawking revealed that black holes evaporate by emitting Hawking radiation \cite{In1}. Furthermore, they possess an entropy proportional to the area of the black hole's event horizon, known as the Hawking-Bekenstein entropy \cite{In2}, and the temperature of black holes corresponds to their surface gravity. By drawing an analogy with ordinary thermodynamic systems, the four laws of black hole thermodynamics were formulated \cite{In3}.
	\\
	
	The study of black holes in anti-de Sitter space (AdS) represents a significant advance in black hole physics, especially with the discovery of the Hawking-Page transition \cite{In4}. Additionally, by considering the cosmological constant as a thermodynamic variable within a framework known as black hole chemistry or extended phase space thermodynamics (EPST), the black hole mass is interpreted as the enthalpy \cite{In5}. In this formalism, the pressure and volume of black holes are introduced as a result of considering the cosmological constant as a thermodynamic variable, where the pressure is related to the cosmological constant by $P = -\Lambda/8\pi G$.
	This formalism has been widely studied and has dramatically enhanced our understanding of black hole physics. Here are some examples: phase transitions of the first and second order \cite{In6,In7}, Joule-Thomson expansion \cite{In8,In9,In10,In11}, the mutual effect between dark energy and black holes \cite{In12,In13}, and multi-critical points \cite{In14}. 
	\\
	
	\par In the AdS/CFT correspondence, the partition function of AdS is equal to that of CFT. In this context, the thermodynamics of AdS black holes is equivalent to the thermodynamics of  CFT  \cite{A144}. In the holographic interpretation, a dictionary relates the quantities of AdS black holes to those of CFT. This thermodynamics is known as holographic thermodynamics. In this framework, there are new thermodynamic variables: the central charge of  CFT, which is related to the AdS radius and the Newton constant, and the chemical potential, which is the conjugate quantity of the central charge \cite{In15}. Moreover, from this point of view, the thermodynamic volume is not the volume of the black hole but the volume of  CFT, which is related to the bulk radius. Here lies the problem: variations in the AdS radius result in non-independent variations of the central charge and the volume of  CFT. To address this problem, there are two approaches: the restricted phase space thermodynamics (RPST) \cite{In16}, where the AdS radius is considered fixed and the Newton constant is variable. The RPST formalism has  been extensively studied  \cite{In17,In18,In19,In20,In21,In22,In23,In24,In25}.\\
	The second approach is the conformal holographic extended thermodynamics (CHET) \cite{In26,In27,In28}. This thermodynamics is based on re-scaling the metric of CFT by a factor independent of the AdS radius, called the conformal factor. As a consequence of this re-scaling, the radius of CFT differs from the AdS radius. This is in contrast to the RPST thermodynamics, where the AdS radius equals the CFT radius. Furthermore, this approach addresses the problem of non-independent variations of the central charge and the volume of CFT without considering the Newton constant as a variable. Instead, the conformal factor is treated as a thermodynamic parameter, and the AdS radius is treated as a variable. This latter point is similar to the EPST formalism for the variation of the AdS radius. In our study, we consider the CHET approach in lower-dimensional systems.
	\\
	
	\par On the other hand, black holes were studied in three dimensions for the first time in 1992 by Banados, Teitelboim, and Zanelli \cite{Banados:1992wn,Banados:1992gq}, known as BTZ black holes (for a review, see ref. \cite{Carlip:1995qv}). Such a kind of black hole has been known as BTZ black hole, and it is a solution to the Einstein field equations in (2+1) dimensions (two spatial dimensions and one time dimension) with a negative cosmological constant, which corresponds to the anti-de Sitter (AdS) space. This significant discovery greatly contributes to enhancing our understanding of low-dimensional gravity \cite{Witten:2007kt}. It has also become a cornerstone for recent developments in gravity, string theory, and AdS/CFT correspondence. In our current paper, we focus on the charged BTZ black holes and their physics in the context of holographic thermodynamics.\\
	
	\par In the aim of investigating the thermodynamic behavior of BTZ black holes, we will consider Tsalis statistics as a framework. This entropy, proposed by Tsallis and Cirto, is a generalization of Boltzmann-Gibbs entropy. In fact, the standard Boltzmann-Gibbs additive theory is not suitable for large-scale gravitational systems with a divergent partition function. Therefore, it must be generalized to a non-additive form, where the entropy of the entire system is not necessarily equal to the sum of the entropies of its subsystems. This generalized form can be applied in all cases \cite{Tsallis:1987eu,Lyra:1997ggy,Tsallis:1998ws}. In our analysis, we specifically examine the effect of the Tsallis parameter, $\delta$, which appears in the Tsallis entropy expression $S_T=S^\delta$ \cite{Tsallis:2012js}, where $S$ is the Hawking-Bekenstein entropy.\\
	
	The motivation for this work is to understand black hole thermodynamics in lower dimensions using the CHET formalism and Tsallis statistics. Focusing on charged BTZ black holes, it aims to clarify their stability and phase transitions compared to the four-dimensional black holes. The study highlights the central charge's influence on thermal evolution and examines how non-extensive statistics impact black hole thermodynamics.
	\\
	
	Our paper is organized as follows: In Sec. \ref{Sec2}, we present the solution of charged BTZ black holes in Anti-de Sitter space-time and their thermodynamic quantities. In Sec. \ ref {Sec3}, we study the thermodynamic behavior using the CHET formalism. In Sec. \ ref {Sec4}, we incorporate Tsallis statistics to examine the effect of non-extensive statistics on the thermodynamic behavior of BTZ black holes. We conclude and discuss our results in Sec. \ ref {Sec5}. In this paper, we adopt the unit $\hbar = c = k_B = 1$.

	\section{BTZ Black Holes}
	\label{Sec2}
	We consider the solution of charged black holes in $(2 + 1)$ dimensions with a negative cosmological constant in Einstein-Maxwell theory. The action of this theory is given by the equation \cite{BTZ1}
	\begin{equation}
		I= \int d^3 x \sqrt{-g}\left(\frac{R - 2 \Lambda}{16 \pi G} - \frac{1}{4} F_{\mu \nu} F^{\mu \nu}\right),
	\end{equation}
	where \(R\) is the Ricci scalar, \(\Lambda\) is the cosmological constant, \(G\) is the gravitational constant, and \(F_{\mu \nu}\) is the electromagnetic field strength tensor. The metric of the charged BTZ black hole in AdS space-time is given by \cite{metric}
	\begin{equation}
		\mathrm{d} s^2 = -f(r) \mathrm{d} t^2 + f(r)^{-1} \mathrm{d} r^2 + r^2 \mathrm{d} \theta^2,
	\end{equation}
	where the function \( f(r) \) is
	\begin{equation}
		f(r) = -8 G M + \frac{r^2}{l^2} - 8 \pi G Q^2 \ln \left(\frac{r}{l}\right),
	\end{equation}
	with \( M \) representing the mass of the black hole, \( Q \) its electric  charge, and \( l \) the AdS radius related to the cosmological constant by \( \Lambda = -1/l^2 \). The mass expression can be obtained by solving $f(r_0)=0$, where $r_0$ is the event horizon. Solving this equation, we find 
	
	\begin{equation}
		\label{mas}
		M=\frac{r_{0}^2}{8 G l^2}-\pi  Q^2 \log \left(\frac{r_{0}}{l}\right).
	\end{equation}
	The Hawking temperature is derived from the metric function as follows 
	\begin{equation}
		\label{T1}
		T=\frac{f'(r_0)}{4\pi} =\frac{r_{0}}{2 \pi l^2}-\frac{2 G Q^2}{r_{0}}.
	\end{equation}
	The charged BTZ black hole cannot evaporate completely due to the conservation of its electrical charge. When the black hole reaches its maximum state of evaporation, it is called an extremal black hole. At this point, the black hole has the lowest possible mass relative to its charge, and its temperature is zero. Using Eq. \eqref{T1}, the event horizon of the extremal black hole can be obtained  
	\begin{equation}
		\label{rmin}
		r_{ext}= 2Ql \sqrt{\pi G}.
	\end{equation}
	The entropy of the BTZ black hole can be obtained using the Hawking-Bekenstein entropy-area law as follows
	\begin{equation}
		\label{S0}
		S=\frac{\pi r_0}{2G}.
	\end{equation}
	For extremal black holes, the entropy $S_{ext}$ is 
	\begin{equation}
		S_{ext}= \frac{\pi^{3/2} Ql}{\sqrt{G}},
	\end{equation}
	The entropy of extremal black holes is proportional to the electric charge and the AdS radius. The extremal entropy is zero for uncharged black holes, indicating a complete evaporation. We derive the electric potential on the event horizon from the mass expression as follows
	\begin{equation}
		\Phi= \dfrac{\partial M}{\partial Q} = -2 \pi Q \ln \left(\frac{r_0}{l} \right).
	\end{equation} 
	\section{Conformal Holographic Extended Thermodynamics}
	\label{Sec3}
	In this section, we study the charged BTZ black holes using the conformal holographic extended thermodynamics (CHET) formalism, where we consider that the AdS radius is different from the CFT radius.
	\subsection{CFT re-scaling}
	To build a thermodynamic framework where the central charge and the volume of CFT vary independently, we need to re-scale CFT in a way that respects its conformal symmetry. This can be expressed through CFT metric as follows \cite{In26}
	
	\begin{equation}
		ds^2 = \omega^2 \left(-dt^2 + l^2 d\Omega_{d-2}\right),
	\end{equation}
	where the thermodynamic parameter, $\omega$, is a dimensionless conformal factor at the AdS boundary. The framework of our study is $AdS_3/CFT_2$ correspondence and CFT  metric within the dimension of the  BTZ black holes, and we obtain  
	\begin{equation}
		ds^2 = \omega^2 \left(-dt^2 + l^2 d\theta\right).
	\end{equation}
	The central charge, \( C \), and  volume, \( \mathcal{V} \), of CFT are expressed by
	
	\begin{equation}
		\label{d0}
		C= \frac{l}{8G} 
	\end{equation}
	and 
	\begin{equation}
		\mathcal{V}= 8\pi\omega l,
	\end{equation}
	respectively. In AdS/CFT correspondence, the thermodynamics of AdS black holes are related to the thermodynamics of the boundary, i.e., CFT thermodynamics. In our work, we relate the thermodynamics of charged BTZ black holes to the thermodynamics of \( CFT_2 \) re-scaling using the following dictionary \cite{In26}
	\begin{equation}
		\label{d1}
		\tilde{S} = S = \frac{A}{4G}, \quad \tilde{E} = \frac{M}{\omega}, \quad \tilde{T} = \frac{T}{\omega}, \quad \tilde{\Phi} = \frac{\Phi \sqrt{G}}{\omega l}, \quad \tilde{Q} = \frac{Q l}{\sqrt{G}}.
	\end{equation}
	where symbols without tildes denote quantities of BTZ black holes, and those with tildes denote the CFT quantities. In this formalism, the black hole mass is proportional to the internal energy by the conformal factor as \( \tilde{E}=M/\omega \). For $\omega = 1$, we recover the Visser's thermodynamic \cite{In15} and restricted phase space thermodynamics \cite{In16}, where the black hole's mass equals the internal energy.

	\subsection{Holographic Thermodynamics}
	We build the Smarr relation and the first law from the AdS/CFT dictionary, Eq. \eqref{d1}, and the on-shell Euclidean action of the gravity theory. Additionally, we obtain a holographic relationship between the state of the black hole and its CFT dual by using the equality of the partition functions of CFT and the gravity theory in AdS, \( Z_{AdS} = Z_{CFT} \) \cite{Z}. We find
	
	\begin{equation}
		\label{F1}
		\mu C = W = T \ln Z_{CFT}= -T \ln Z_{AdS}  = T I_E,
	\end{equation}
	where \(\mu\) is the chemical potential and \(W\) is the free energy. In this paper, we study charged BTZ black holes in 3D within the framework of Einstein-Maxwell theory. The on-shell Euclidean action is expressed in terms of black hole quantities as follows \cite{Z1}
	
	\begin{equation}
		\label{F2}
		I_E = \beta \left(M - TS - \Phi Q \right),
	\end{equation}
	where $\beta= 1/T$ is the inverse of black hole temperature. From Eqs. \eqref{F1} and \eqref{F2}, we find 
	\begin{equation}
		\label{F3}
		M= TS + \Phi Q + \mu C.
	\end{equation}
	From Eqs \eqref{d1} and \eqref{F3}, we find the Smarr relation in CHET formalism of charged BTZ black holes, as flows
	\begin{equation}
		\label{F4}
		\tilde{E}= \tilde{T} S+ \tilde{\Phi} \tilde{Q} + \tilde{\mu} C,
	\end{equation}  
	where \(\tilde{\mu}= \mu/\omega\). To obtain the first law, we use the following setup
	\begin{equation}
		\label{F5}
		d\tilde{E}= \frac{d M}{\omega} - \frac{M}{\omega^2} d\omega,
	\end{equation}
	where, $d M$  can be expressed as \cite{In15}
	\begin{equation}
		\label{M11}
		d M= TdS+ \frac{\Phi}{l} d\left(Ql \right)  - M\frac{dl}{l} + \left( M-TS-\Phi Q \right) \dfrac{d \left(l/G \right) }{\left(l/G \right) }.
	\end{equation}
	From Eqs. \eqref{F4}-\eqref{M11}, the first law can be expressed as follows
	\begin{equation}
		d\left(\frac{M}{\omega}\right)=  \frac{T}{\omega} dS+\frac{\Phi \sqrt{G}}{\omega l} d\left(\frac{Q l}{\sqrt{G}}\right) 
		+\left(\frac{M}{\omega}-\frac{T S}{\omega}-\frac{\Phi Q}{\omega}\right) \frac{d\left(l / G\right)}{l / G} 
		-M \frac{d(\omega l)}{\omega^2 l} .
	\end{equation}
	We can simplify the first law using the dictionary, Eq. \eqref{d1},  the expression of the central charge, and the CFT volume in Eq. \eqref{d0}. 	We find
	
	\begin{equation}
		\label{d2}
		d \tilde{E}=\tilde{T} d S+ \tilde{\Phi} d \tilde{Q}+\tilde{\mu} d C-p d \mathcal{V},
	\end{equation}
	Here,   the boundary pressure, $p$, is related to the energy and CFT volume by the following  relationship
	\begin{equation}
		\label{pve}
		\tilde{E}=p\mathcal{V}.
	\end{equation}
	The pressure in the holographic first law is not dual to the bulk pressure, $P$, used in the formalism of extended black holes thermodynamics, which it is relates to the cosmological constant $\Lambda$ by,  $P=- \Lambda/ 8\pi G$. To remove the term $-p\mathcal{V}$, we re-scale the thermodynamic quantities as follows
	\begin{equation}
		\bar{M}=\bar{E}= \omega l\tilde{E}, \qquad \bar{T}= \omega l \tilde{T}, \qquad \bar{\Phi} = \omega l \tilde{\Phi}, \qquad \bar{\mu}= \omega l \tilde{\mu}.
	\end{equation} 
	Finally, we find the expression of the first law and Smarr relation in the CHET formalism for charged BTZ black holes as follows
	\begin{equation}
		d \bar{M} = \bar{T}dS+  \bar{\Phi} d \tilde{Q}   + \bar{\mu} dC
	\end{equation}
	and 
	
	\begin{equation}
		\label{Sm}
		\bar{M} = \bar{T}S+  \bar{\Phi} \tilde{Q}   + \bar{\mu} C,
	\end{equation}
	respectively. This Smarr relation in CHET for BTZ black holes exhibits first-order homogeneity. Additionally, we re-scale the extensive parameters. Also, we  re-scale the extensive parameter, $S,\, \tilde{Q}$ and $C$,    by $S \rightarrow \lambda S, \tilde{Q} \rightarrow \lambda \tilde{Q}, C \rightarrow \lambda C$. Under this re-scaling, $\bar{M}$ remains first-order homogeneous. This property of homogeneity is preserved in lower dimensions. \\
	
	After constructing the first law and the Smarr relation, we study the thermodynamics behavior of charged BTZ black holes in CHET formalism. Firstly, we find the thermodynamics quantities of these black holes in CHET formalism by use of the first law and the expression of black hole mass 
	\begin{equation}
		\label{YT}
		\bar{T}= \left( \dfrac{\partial \bar{M}}{\partial S}\right) _{\tilde{Q},C}= \frac{S^2-\pi ^3 \tilde{Q}^2}{8 \pi ^2 C S},
	\end{equation}    
	\begin{equation}
		\label{YP}
		\bar{\Phi} = \left( \dfrac{\partial \bar{M}}{\partial \tilde{Q}}\right) _{S,C}= -\frac{\pi  \tilde{Q} \log \left(\frac{S}{4 \pi  C}\right)}{4 C},
	\end{equation}
	and
	\begin{equation}
		\label{YM}
		\bar{\mu} = \left( \dfrac{\partial \bar{M}}{\partial C}\right) _{S,\tilde{Q}}= \frac{2 \pi ^3 \tilde{Q}^2 \log \left(\frac{S}{4 \pi  C}\right)+2 \pi ^3 \tilde{Q}^2-S^2}{16 \pi ^2 C^2}.
	\end{equation}
	The following expression can express the heat capacity
	
	\begin{equation}
		\label{YC}
		\zeta= \bar{T} \left(  \dfrac{\partial S}{\partial \bar{T}}\right)_{\tilde{Q},C}= \frac{S \left(S^2-\pi ^3 \tilde{Q}^2\right)}{\pi ^3 \tilde{Q}^2+S^2}. 
	\end{equation}
We observe that the heat capacity,  Eq. \eqref{YC}, is always positive because the black hole entropy is always greater than or equal to the extremal entropy, $S_{ext} = \pi^{3/2} \tilde{Q}$. This indicates that BTZ black holes are always stable.
	\begin{figure}[htp]
		\centering
		\includegraphics[width=0.7\linewidth]{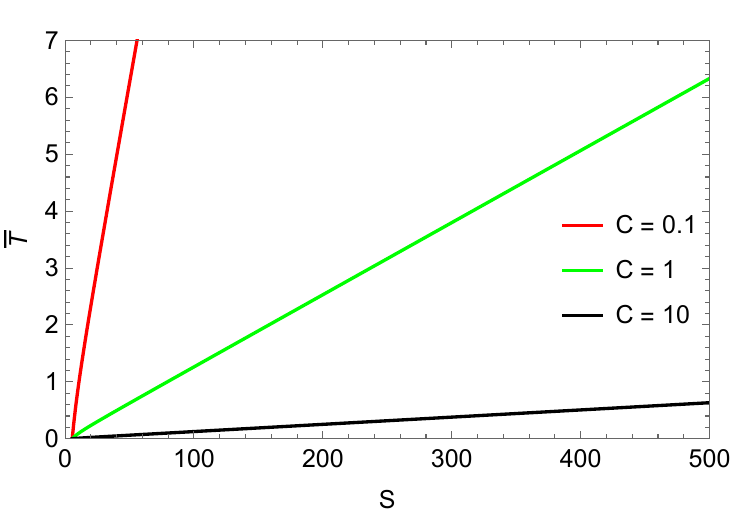}
		\caption{ $\bar{T} - S$  curve for different values of central charge, with the electric charge fixed at $ \tilde{Q}=1$}
		
		\label{Fig1}
	\end{figure}
	
	Fig. \ ref {Fig1} represents the thermal evolution of charged BTZ black holes in the thermodynamic process where the electric charge is fixed for different values of the central charge. This figure shows that the thermodynamic behavior is always in one phase for both small and large black holes. Additionally, we observe that the form of thermal evolution is proportional to the black hole's entropy, and the inclination depends on the inverse value of the central charge. From Eq. \eqref{YT}, we make an approximation to express this relation. For large entropy, we find
	\begin{equation}
		\label{YT1}
		\bar{T} = \alpha S
	\end{equation}
	where $\alpha = 1/ 8\pi^2 C$. We conclude that the electric charge of the BTZ black hole does not affect the thermodynamic evolution of large black holes. Still, it does affect the value of the minimum entropy or the entropy of extremal black holes. Also, the central charge of CFT affects thermal evolution, with a small central charge leading to a significantly larger acceleration of evolution than a large central charge.   
	\\
	
	Unlike four-dimensional charged black holes \cite{In6}, which undergo- and second-order phase transitions and exhibit critical phenomena, three-dimensional charged black holes are always stable and do not experience these phase transitions. This stability can be derived from Fig. \ref{Fig1}, Eq. \eqref{YT}, and the heat capacity expression in Eq. \eqref{YC}, which is always positive. Consequently, the two-dimensional CFT dual of BTZ black holes is always stable and maintains consistent thermodynamic behavior regardless of thermodynamic quantities. In contrast, the CFT dual of a four-dimensional black hole shows thermodynamic behavior dependent on thermal quantities, resulting in phases of stability and instability.

	\section{BTZ Black Holes with Tsallis Statistics}
	\label{Sec4}
	In this section, we study charged BTZ black holes using the CHET formalism with corrections based on Tsallis statistics. We investigate the thermodynamic behavior of BTZ black holes and their CFT duals, focusing on stability, phase transitions, and critical phenomena within the context of Tsallis statistics.
	\subsection{Tsallis entropy}
In this subsection, we present a modified version of the Conformal Holographic Extended Thermodynamics formalism using Tsallis statistics \cite{Z2}. Our motivation for incorporating Tsallis statistics into holographic thermodynamics stems from holographic duality. Specifically, for a 4-dimensional black hole, the Boltzmann-Gibbs entropy is proportional to its surface area, \( L^2 \), where \( L \) is a characteristic linear scale. More generally,  if the number of dimensions is equal to d=2, the entropy scales with ln L, while for $d > 2$, it is proportional to $L^{d-2}$. These results illustrate the non-extensive nature of entropy in systems with holographic descriptions across \( d \) dimensions. In contrast, systems that adhere to Gibbs-Boltzmann statistics exhibit extensive entropy, where combining two systems results in an additive entropy proportional to the system's volume. Systems with a holographic description deviate from this property, necessitating a non-extensive statistical framework, such as Tsallis statistics, which provides an effective tool for this analysis.

 The Tsallis entropy of black holes can be expressed in terms of the Hawking-Bekenstein entropy, \( S \), and the Tsallis parameter, \(\delta\), as follows  \cite{Tsallis:2012js}
	
	\begin{equation}
		\label{ST}
		S_T = S^\delta.
	\end{equation}
	We recover the Hawking-Bekenstein entropy for the specific value of the Tsallis parameter,   $\delta =1$. Using Eqs. \eqref{S0} and \eqref{ST}, we find the Tsallis entropy of charged BTZ black holes
	\begin{equation}
		\label{SB}
		S_T= \left( \frac{\pi r_0}{2G}\right)^{\delta}.
	\end{equation}
	After adjusting the black hole entropy using Tsallis statistics, we observe that the entropy is no longer directly proportional to the event horizon's area. This behavior closely resembles Barrow's entropy, arising from the deformation of the event horizon under the influence of quantum gravity effects \cite{Br}.
	\\
	
	We determine the Tsallis temperature of the BTZ black hole by employing the Tsallis entropy, Eq. \eqref{SB} and the black hole mass, Eq. \eqref{mas},
	\begin{equation}
		T_T= \dfrac{\partial M}{\partial S_T}= \frac{1}{\delta} \left(\frac{\pi r_0}{2G}\right)^{1-\delta } \left(\frac{r_0}{2 \pi  l^2}-\frac{2 G Q^2}{r_0}\right).
	\end{equation}
	We construct the Tsallis correction of the CHET formalism by incorporating the entropy and temperature of Tsallis. We modify the first law and Smarr relation accordingly. We establish a relationship between the standard quantities $(S,T)$ and Tsallis quantities $(S_T, T_T)$ as follows
	
	\begin{equation}
		\label{STST}
		TS= \delta\, T_T S_T.
	\end{equation}
	From Eqs. \eqref{Sm} and \eqref{STST}, we deduce the Tsallis correction of the Smarr relation as follows
	\begin{equation}
		\bar{M} = \delta\, \bar{T}_T S_T+  \bar{\Phi} \tilde{Q}   + \bar{\mu} C,
	\end{equation}
	In Tsallis statistics, \( \bar{M} \) is a homogeneous function of \((\bar{T}_T, S_T)\) of order \(\delta\), and of \((\bar{\Phi}, \tilde{Q})\) and \((\mu, C)\) of order 1. This finding contradicts the typical Smarr relation in CHET formalism in the normal case, i.e., without corrections, where \( M \) is uniformly a homogeneous function of order 1 for all variables (see Eq. \eqref{Sm}). Thus, the homogeneity of the Smarr relation in CHET thermodynamics is violated in Tsallis statistics. We  derive the first law and the Gibbs-Duhem equation from the Smarr relation by
	
	\begin{equation}
		d\bar{M}= \bar{T}_T dS_T+ \bar{\Phi} d \tilde{Q} + \bar{\mu} d C,
	\end{equation} 
	
	\begin{equation}
		d\bar{\mu}=- \delta \mathcal{S}_T d \bar{T} - \mathcal{\tilde{Q}}d \bar{\Phi},
	\end{equation} 
	respectively, where $\mathcal{S}_T = S_T / C$ and $\mathcal{\tilde{Q}} = \tilde{Q} / C$. To study the effect of the Tsallis parameter on thermodynamic behaviors, we write the new thermodynamic quantities in Tsallis statistics. We obtain these quantities from the first law as follows
	\begin{equation}
		\label{Te1}
		\bar{T}_T=  \left( \dfrac{\partial \bar{M}}{\partial S_T}\right) _{\tilde{Q},C}=\frac{S_T^{2/\delta} - \pi^3 \tilde{Q}^2}{8 \pi^2 \delta C  S_T},
	\end{equation}
	\begin{equation}
		\bar{\Phi} = \left( \dfrac{\partial \bar{M}}{\partial \tilde{Q}}\right) _{S_T,C}= -\frac{\pi \tilde{Q}}{4C} \log{\left(\frac{S_T^{1/\delta}}{4  \pi C}\right)}, 
	\end{equation}
	and
	\begin{equation}
		\bar{\mu} = \left( \dfrac{\partial \bar{M}}{\partial C}\right) _{S_T,\tilde{Q}}= -\frac{S_T^{2/\delta} - 2 \pi^3 \tilde{Q}^2 \left(\log{\left(\frac{S_T^{1/\delta}}{4 \pi C}\right)} + 1\right)}{16 \pi^2 C^2}.
	\end{equation}
	
	\subsection{Phase Structure}
	The thermodynamic behavior of black holes depends on various parameters, including mass, charge, and rotational momentum. Additionally, the thermodynamic behavior is influenced by the fields surrounding the black hole. For instance, the thermodynamic behavior of black holes surrounded by quintessence dark energy is different from that of AdS black holes. Furthermore, the number of dimensions affects the thermodynamic behavior. In our present work, the thermodynamic behavior and evolution of charged black holes in 3D are radically different from their counterparts in 4D, such as Reissner-Nordström AdS black holes.
	\\
	
	Now, we study the effect of the statistical approach on thermodynamic behavior. Specifically, we examine the phase structure and stability of BTZ black holes and their CFT dual in the CHET formalism using Tsallis statistics.
	\\
	
	The first thermodynamic characteristic of BTZ black holes in this approach is the absence of critical phenomena, i.e., second-order phase transitions do not occur. We conclude that this absence of second-order phase transitions is a universal property of BTZ black holes. However, a first-order phase transition can occur under special conditions. We derive the expression for the phase transition entropy, $S_{0}$ , by solving the following equation
	\begin{equation}
 \label{AAA}
		\left(\dfrac{\partial T_T}{\partial S_T} \right)_{\tilde{Q},C}=0 , 
	\end{equation}      
	we find 
	\begin{equation}
		\label{Se1}
		S_{0}= \left(\frac{\pi^{3/2} \tilde{Q} \sqrt{\delta}}{\sqrt{\delta-2}} \right)^\delta.
	\end{equation}
	The entropy of phase transition is related to the value of Tsallis parameter. The values entropy of phase transition are real only when $\delta>2$. The transition temperature $\bar{T}_{0}$ can be obtained from Eqs. \eqref{Te1} and \eqref{Se1} as follows
	\begin{equation}
		\label{Tpt}
		\bar{T}_{0}= \frac{\pi^{1 - \frac{3 \delta}{2}} \tilde{Q}^2 }{4C (-2 + \delta) \delta} \left(\frac{\sqrt{-2 + \delta}}{\tilde{Q} \sqrt{\delta}}\right)^\delta.
	\end{equation}
	To normalize the entropy and temperature, we utilize the phase transition quantities
	\begin{equation}
		s= \frac{S}{S_{0}}, \qquad t=\frac{\bar{T}}{\bar{T}_{0}}
	\end{equation}
	Through this normalization, the central charge of CFT is eliminated, and therefore, it will not affect the thermodynamic behavior. However, it will affect the scale. In other words, it will not affect the occurrence of phase transitions and stability, but it will affect the values of non-normalized thermodynamic quantities.
	\\
	
	To study the stability of black holes, we use the heat capacity,  $\zeta$. 
	The expression for the heat capacity can be obtained as follows.
	
	\begin{equation}
		\label{HC}
		\zeta = \bar{T}_T \left( \frac{\partial S_T}{\partial \bar{T}_T}\right)_{\tilde{Q},C} = -\frac{\delta S \left(S^{2/\delta }-\pi ^3 Q^2\right)}{(\delta -2) S^{2/\delta }-\pi ^3 \delta Q^2}.
	\end{equation}
	
	\begin{figure}[htp]
		\centering
		\includegraphics[width=0.47\linewidth]{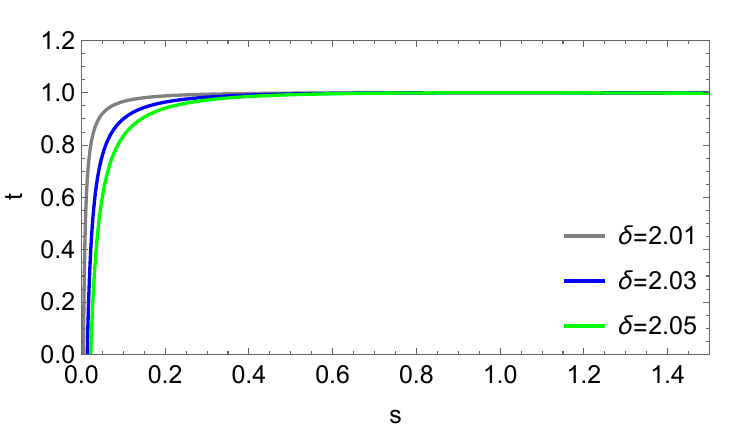}
		\includegraphics[width=0.47\linewidth]{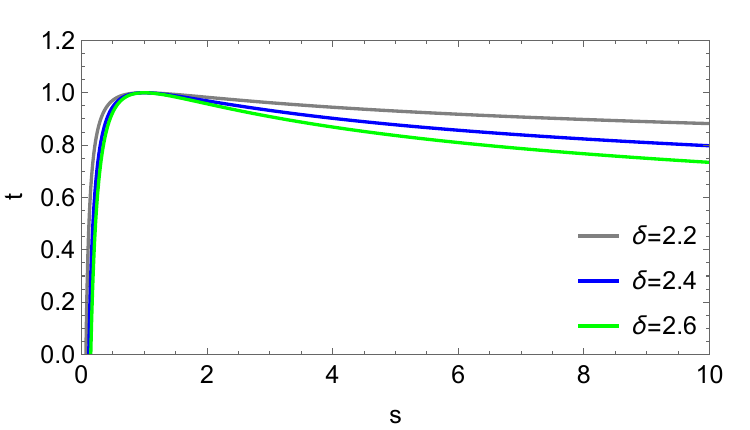}	
		\caption{t-s curves for different values of the Tsallis parameter, with $\tilde{Q}=1$ }
		\label{fig2}
	\end{figure}
\begin{figure}[htp]
		\centering
		\includegraphics[width=0.4\linewidth]{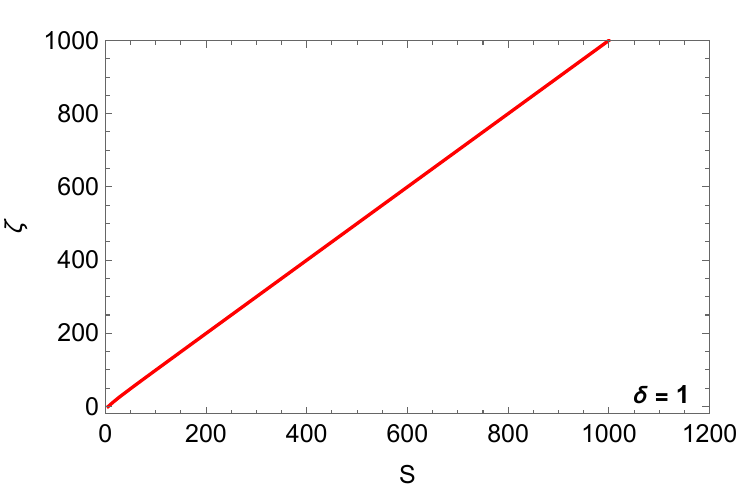}
			\includegraphics[width=0.42\linewidth]{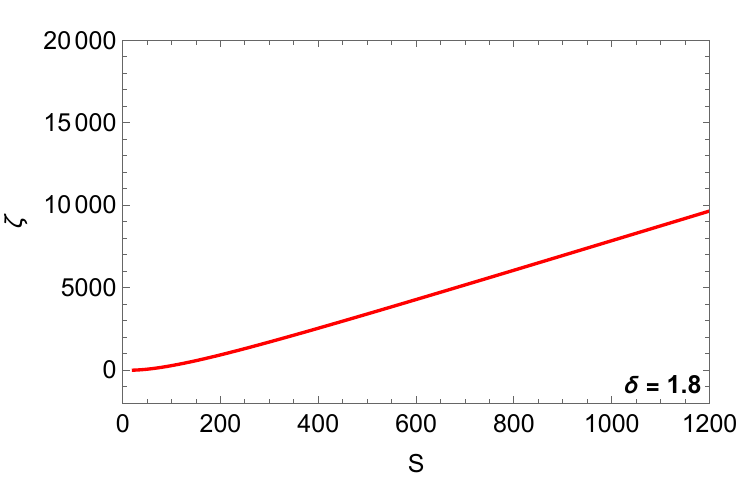}
				\includegraphics[width=0.43\linewidth]{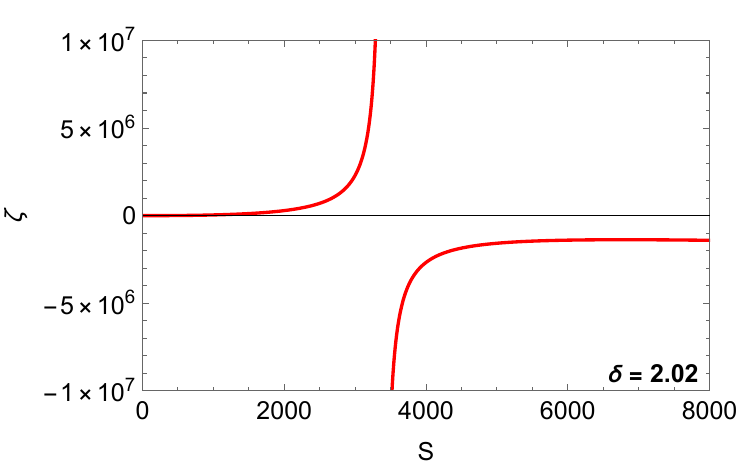}
					\includegraphics[width=0.432\linewidth]{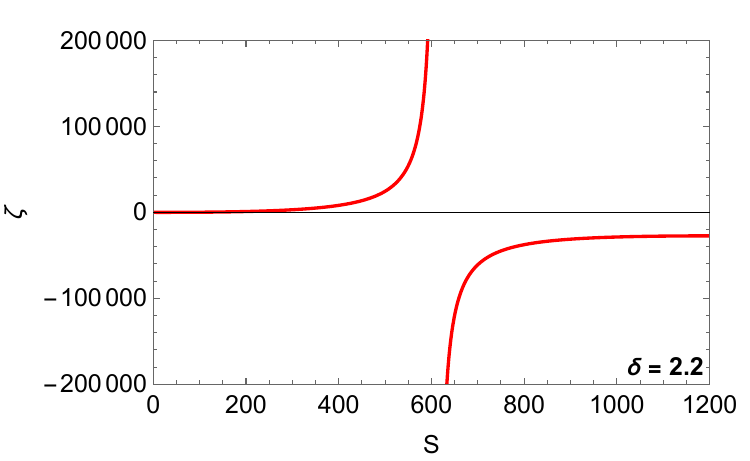}
		\caption{Heat capacity in terms of Tsallis entropy for various values of \(\delta \), with \(\tilde{Q} = 1\).}
		\label{fig3}
	\end{figure}

Fig. \ref{fig2} shows the evolution of the Tsallis temperature as a function of the Tsallis entropy for different values of the Tsallis parameter \(\delta\), with the electric charge fixed at \(\tilde{Q} = 1\). From this figure and Eq. \eqref{Se1}, we observe that for \(\delta > 2\), a first-order phase transition occurs between small and large black holes. In contrast, for \(\delta < 2\), there is no real solution to Eq. \eqref{AAA}, indicating that no phase transition occurs.	Additionally, in Fig. \ref{fig3}, we observe two distinct regimes for the heat capacity based on the value of \(\delta\). In the first case, when \(\delta > 2\), small BTZ black holes are stable (corresponding to positive heat capacity), while large BTZ black holes are unstable (corresponding to negative heat capacity). In the second case, when \(\delta < 2\), the heat capacity is positive, indicating that BTZ black holes are always stable.
	\\
	
	For most studies of the Tsallis statistics on black hole thermodynamics, we introduce the Helmholtz free energy for BTZ black holes in the CHET formalism as follows:
	
	\begin{equation}
		\bar{F}= \bar{M} - \bar{T}_T S_T,
	\end{equation}
	
	We can normalize the Helmholtz free energy as \( f = \bar{F}/\bar{F}_{TP} \), where \( \bar{F}_{TP} \) corresponds to the free energy at the phase transition.
	\\
	
	\begin{figure}[htp]
		\centering
		\includegraphics[width=0.4\linewidth]{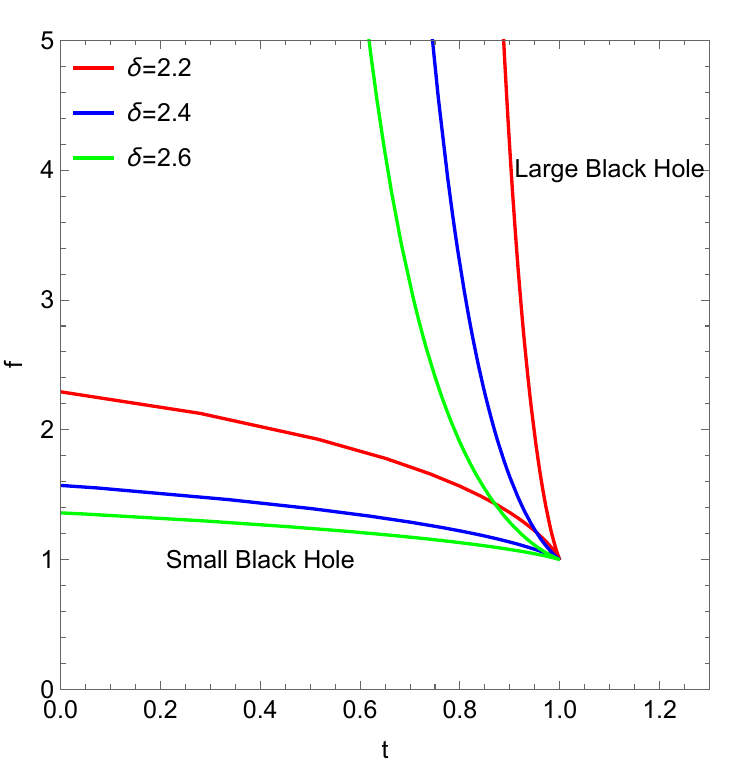}
		\includegraphics[width=0.412\linewidth]{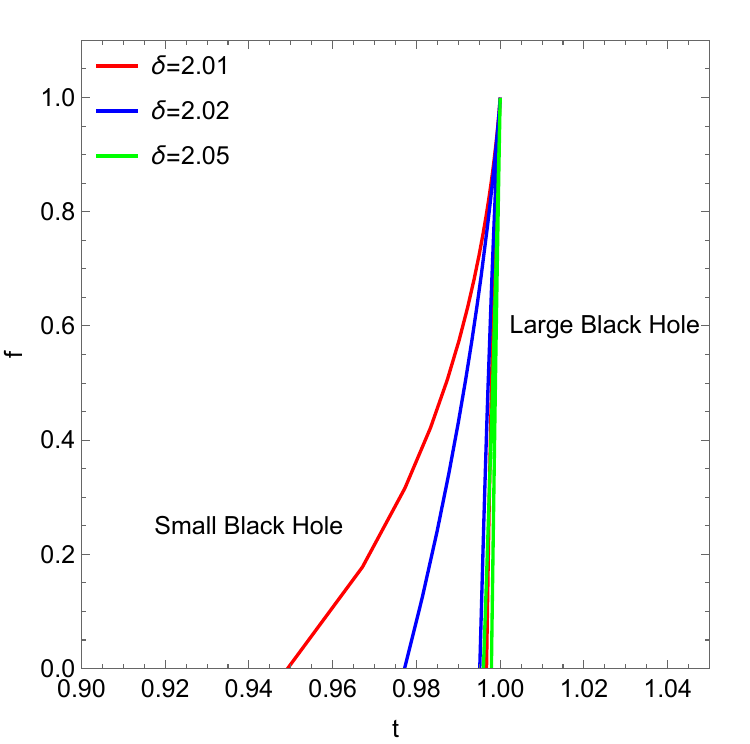}
		\caption{\( f \)-\( t \) curves for different values of the Tsallis parameter, with \( \tilde{Q} = 1 \).}
		\label{fig4}
	\end{figure}
	
	We plot in Fig. \ref{fig4} the Helmholtz free energy as a function of the Tsallis temperature for two cases of the Tsallis parameter. In the first case, the Tsallis parameter is far from 2, and in the second case, \( \delta \approx 2 \). In both cases, we observe two branches: a branch of small black holes and a branch of large black holes. The effect of Tsallis statistics is evident in the curves of the Helmholtz free energy, as the behavior in the first case is significantly different from that in the second case. In the first case, the Helmholtz free energy decreases with increasing temperature for both the small and large black hole branches. The free energy must also be greater than or equal to the Helmholtz free energy corresponding to the phase transition. In the second case, the Helmholtz free energy increases with increasing temperature, and the Helmholtz free energy is less than or equal to the Helmholtz free energy corresponding to the phase transition.

	\section{Discussions and Conclusions}
	\label{Sec5}

	This paper provides a comprehensive analysis of the conformal holographic extended thermodynamics in lower dimensions, specifically focusing on charged BTZ black holes. It also examines the effect of non-extensive statistics on thermodynamic behavior.
	\\
	
	Firstly, we present the solution for charged BTZ black holes in Einstein-Maxwell theory in three dimensions with a negative cosmological constant, i.e., in Anti-de Sitter space-time. We also derive the expressions for the black hole's mass, Hawking temperature, and Bekenstein-Hawking entropy. Additionally, we determine the minimal entropy of charged BTZ black holes, which corresponds to the entropy of extremal black holes.
	\\
	
	Secondly, we established a version of the CHET formalism for Einstein-Maxwell theory in three dimensions, based on the holographic interpretation where the thermodynamics of AdS black holes is equivalent to that of CFT dual. This approach involves re-scaling the CFT metric by a conformal factor and considering this factor as a thermodynamic parameter. Unlike the RPST formalism, where the Newton constant is variable, and the AdS radius is fixed,  the Newton constant is fixed, and the AdS radius is variable. We also presented a holographic dictionary that relates AdS black hole quantities to the  CFT quantities.
	
	We then examined the thermodynamic behavior of charged BTZ black holes and their CFT dual. We derived the first law of thermodynamics and the Smarr relation within the CHET formalism. The Smarr relation obtained in the CHET formalism for BTZ black holes maintains first-order homogeneity. Our study revealed that charged BTZ black holes exhibit stable thermodynamic behavior without undergoing first or second-order phase transitions, unlike their four-dimensional counterparts. This stability is confirmed by the positive heat capacity derived from the thermodynamic quantities. For large black holes, the thermal evolution is directly proportional to their entropy, and the rate of this evolution is inversely related to the central charge. Smaller central charges of CFT result in a more significant acceleration of the thermal evolution.
	\\

	Lastly, by incorporating Tsallis statistics, which generalize Boltzmann-Gibbs statistics for non-extensive systems, we observed modifications in the thermodynamic properties of BTZ black holes. In the framework of Tsallis statistics, the thermodynamic behavior directly depends on the Tsallis parameter.   For the case where \(\delta < 2\), the black holes are always stable; i.e., no phase transition occurs. However, for the case where \(\delta > 2\), we found that a first-order phase transition occurs between small and large black holes, with small BTZ black holes being stable while large black holes are unstable. This correction highlights the importance and influence of alternative statistical frameworks in black hole thermodynamics. Nevertheless, the microscopic nature of Tsallis entropy remains unclear and poses a significant challenge in theoretical physics. While some studies investigate the microscopic origins of black hole entropy in string theory for specific cases \cite{MN1,MN2,MN3}, these efforts are limited in scope. As a result, the microscopic account of Tsallis entropy for BTZ black holes also remains unresolved.
	\\

	In summary, this paper advances our understanding of the holographic thermodynamics of black holes in lower dimensions by studying charged BTZ black holes within the CHET formalism and exploring the implications of Tsallis statistics. The findings underscore the critical role of the central charge and non-extensive statistics in shaping the thermodynamic behavior of BTZ black holes. These insights contribute to the broader pursuit of understanding black hole thermodynamics and the AdS/CFT correspondence.
	
\section*{Acknowledgments}

Y. Ladghami would like to express gratitude for the support received from the "PhD-Associate Scholarship – PASS" grant (number 1019UMP2023) provided by the National Center for Scientific and Technical Research in Morocco.

\end{document}